\documentclass[10pt,a4paper]{article}
\usepackage{cite}

\usepackage{graphicx}

\usepackage{color}

\def\lf{l_{\rm f}}

\def\Es{{E_{\rm s}}}
\def\Eb{{E_{\rm b}}}
\def\Ec{{E_{\rm c}}}

\def\tlag{{t_{\rm lag}}}

\def\alphaeff{\alpha_{\rm eff}}
\def\ne{n_{\rm e}}
\def\nes{n_{\rm es}}
\def\neb{n_{\rm eb}}
\def\nec{n_{\rm ec}}
\def\we{w_{\rm e}}
\def\wes{w_{\rm es}}
\def\web{w_{\rm eb}}
\def\wec{w_{\rm ec}}

\def\vs{v_{\rm s}}
\def\rs{r_{\rm s}}
\def\vec#1{{\bf #1}}

\def\citep#1{\cite{#1}}
\def\d{{\rm d}}

\begin{document}

\title{Sub-nanosecond delays of light emitted by streamer in atmospheric pressure air: 
analysis of N$_2$(C$^3\Pi_u$) and N$_2^+($B$^2\Sigma_u^{+}$) emissions and fundamental streamer structure}


\author{{\bf Tom\'a\v s Hoder}\\
Leibniz Institute for Plasma Science \\ and Technology (INP Greifswald),
\\ Felix-Hausdorf-Str.2,17489 Greifswald, Germany \\ \\
Department of Physical Electronics,\\ Faculty of
Science, Masaryk University, \\ Kotl\'a\v rsk\'a 2,
611 37 Brno, Czech Republic 
\\
\texttt{hoder@physics.muni.cz}\\ \\
\and
{\bf Zden\v ek Bonaventura}\\
Department of Physical Electronics,\\ Faculty of
Science, Masaryk University, \\ Kotl\'a\v rsk\'a 2,
611 37 Brno, Czech Republic \\
\texttt{zbona@physics.muni.cz} \\ \\
\and
{\bf Anne Bourdon}\\
CNRS, UPR 288 Laboratoire EM2C, \\
Grande voie des vignes, 92295 Ch\^atenay-Malabry, France\\ \\
Ecole Centrale Paris, Grande voie des vignes, \\
92295 Ch\^atenay-Malabry, France\\
\texttt{anne.bourdon@ecp.fr} \\ \\
\and
{\bf Milan \v Simek}\\
Department of Pulse Plasma Systems, Institute of Plasma Physics,\\
Academy of Sciences of the Czech Republic,\\
Za Slovankou 3, 182 00 Prague, Czech Republic\\
\texttt{simek@ipp.cas.cz}
}

\maketitle
\thispagestyle{empty}

\vbox{
\begin{abstract}
Both experimental and theoretical analysis of an {ultra-short} phenomena occurring during the positive streamer propagation in atmospheric pressure air is presented. 
{With tens-of-picoseconds and tens-of-microns precision, it is shown that when the streamer head passes a spatial coordinate, emission maxima from N$_2$ and N$_2^+$ radiative states follow with different delays. 
These different delays 
are caused by differences in the dynamics of populating the radiative states, 
due to different excitation and quenching rates.} 
{Associating the position of the streamer head with the maximum value of the self-enhanced electric field, a delay of 160$\,$ps was experimentally found for the peak emission of the first negative system of N$_2^+$.}
{For the first time, a delay dilatation was observed experimentally on early-stage streamers and clarified theoretically.}
{Thus, in the case of second positive system of N$_2$ the delay can reach as much as 400$\,$ps.}
In contrast to the highly-nonlinear behaviour of streamer events, it is shown theoretically that emission maxima delays linearly depend on the ratio of the streamer radius and its velocity. 
This is found to be one of the fundamental streamer features and its use in streamer head diagnostics is proposed.
Moreover, radially-resolved spectra are synthetized for selected subsequent picosecond moments in order to visualize spectrometric fingerprints of radial structures of N$_2$(C$^3\Pi_u$) and N$_2^+$(B$^2\Sigma_u^+$)  populations created by streamer-head electrons.
\end{abstract}
}

\section{Introduction}
Streamer in atmospheric pressure air is a contracted ionizing wave 
that propagates into a low- or non-ionised medium exposed to a high electric field. 
It is characterised by a self-generated field enhancement at the head of the growing discharge channel, leaving  a trail of filamentary plasma behind. Such a wave phenomenon results from the space charge left by electron avalanches \cite{marode,ebert2006}.
Streamers are present in a large number of plasmas, whether operated in the laboratory \cite{stari2012,briels,akishev,chen,huiskamp,nudnova}, in industrial applications \cite{cernak2011} or occurring in lightning and transient luminous events in upper atmosphere \cite{ebert2010,kanmae,luque,pasko2007,neubert}.

In recent time, 
raising interest to investigate this ultra-fast phenomenon has been enabled by better accessibility of fast gated intensified CCD cameras. 
However, high-speed camera investigations neglect one very important fact: increasing the temporal resolution of measuring devices to nanoseconds (or a bit under) is  insufficient to follow the basic processes within the streamer discharge in atmospheric pressure air.
As a consequence, one has to take into account additional effects. 
One of them is, e.g., the influence of nanosecond gated recordings on the accuracy of the electric field strength estimation from the ratio of emission intensities of the (0,0) vibrational bands of second positive system of N$_2$ (SPS, with spectral band head at 337.1$\,$nm) and first negative system of N$_2^+$ (FNS, at 391.5$\,$nm)  
by ICCD cameras, as shown in \cite{naidis,bonaventura2011,celestin2010}. 

Typically, the estimation of basic parameters of practically all nitrogen-containing plasmas at different pressures is widely based on the emission of the two above mentioned nitrogen spectral systems  \cite{paris2005,bibinov,jelenkovic1987,stanfield,omholt} - dominantly due to the large difference in their excitation potentials.
Thus, also for streamers in atmospheric pressure air, these emissions have been in focus for a long time as well  \cite{naidis,bonaventura2011,galimberti,hartmann,marode1975,ikuta1976,ikuta1990,creyghton1994,kozlov2001}.
Mutual delay (or shift) of the SPS and FNS emission signal maxima (i.e. SPS-to-FNS delay) of propagating streamer 
can be found in older literature dealing with time and space highly-resolved streamer investigations. 
In 1976, Ikuta and Kondo \cite{ikuta1976} applied probably for the first time the time-correlated single photon counting (TC-SPC) based technique in the investigation of streamer discharges. 
From their results this delay is visible. 
Moreover, in \cite{creyghton1994} it is shown theoretically that FNS and SPS emission maxima 
occur with different delays behind the electric field peak, but not commented.
From the theoretical works of Wang et al. \cite{wang1990} or Kulikovsky \cite{kulikovsky1998} one can learn about the synchronised development of the effective ionization (or excitation) rates in comparison with the electrical field and electron density development. 
Nevertheless, its detailed impact on the spectrally resolved emission development has not been discussed yet.
The presence of  SPS-to-FNS delay is analysed in detailed works of Matveev, Djakov and co-workers \cite{matveev,djakov} based on the 1D simulations of Djakov et al. \cite{djakov1}. 
In their papers, the authors studied theoretically the influence of the spatial and temporal resolution on the determination of the electric field by possible experimental approach.
For cases under investigation one can conclude that the temporal resolution of the spectrometric device is a very important parameter which should be of order of tens of picoseconds and the spatial resolution of few tens of microns. 
Only under such conditions the determination of the electric field by the FNS/SPS intensity ratio method is not distorted significantly \cite{djakov,shcherbakov1997}. 
Shcherbakov and Sigmond \cite{shcherbakov2007,shcherbakov2007b} applied the TC-SPC technique  with sufficiently high resolution and emphasised the necessity to have a high enough temporal resolution to be able to resolve SPS-to-FNS delay. 
Obviously, if this is not the case the estimation of the synchronous electric field or even exact peak field values  by the ratio method fails. 

Recently, this approach was further theoretically developed by Naidis in \cite{naidis}. Bonaventura, Celestin and co-workers \cite{bonaventura2011,celestin2010} analysed theoretically the effect of these delays on ICCD-based electric field estimation in more details (also applying this to streamers in sprites in upper atmosphere in \cite{celestin2010}) adding the influence of the radial integration over the streamer structure which was not possible in the 1.5D model of Naidis \cite{naidis} or in the { 0D }pioneering work of Djakov \cite{djakov1}.

Even though there is an increased attention to this topic in last years, 
so far, 
{no study has investigated in detail the emission maxima peculiarities with all consequences for streamer spectroscopy and diagnostics.} 
In the present work, motivated by high-precision experimental results, systematic computer simulations were performed to study in detail emission delays.
The study is based on well established 2D streamer model which enables deeper insight into the delay-issue than previous 0D or 1.5D approaches or computations in 2D focused only on nanosecond-gated cameras. 
Based on the model of Kulikovsky \cite{kulikovsky1998} and our results from 2D computer simulations, analytical description for the delay-parameter is developed.

The manuscript is organised as follows. Section 2 summarizes  motivations for the present work based on recent experimental findings and the strategy proposed for the modelling.
In section 3 the 2D streamer model is described. 
The section 4 analyses numerical outcomes. The origin of delays is discussed in detail and its analytical expression and linear dependency are presented in subsections 4.1 and 4.2, respectively. Finally, in the last subsection 4.3, the high-resolved synthetic spectral representation of the streamer head is presented, enabling detailed insight into the radial structure of the streamer-head emission.

{
\section{Experimental motivation and strategy adopted for the modelling}

In this work, we have carefully analysed experimental data obtained in two different streamer-discharge setups.
We applied the TC-SPC technique on the negative corona Trichel pulses \cite{hoder2012} and barrier discharges \cite{hoder2010} both in atmospheric pressure air and the synchronous developments of the SPS and FNS emissions for positive streamers were recorded and analysed. Their ratio was computed and according to the simple kinetic scheme \cite{paris2005,kozlov2001,hoder2012} the corresponding electric field development was determined for a selected coordinate of tens of micrometers dimension.
}

\begin{figure}[h!]
\begin{center}
\includegraphics[width=0.3\columnwidth]{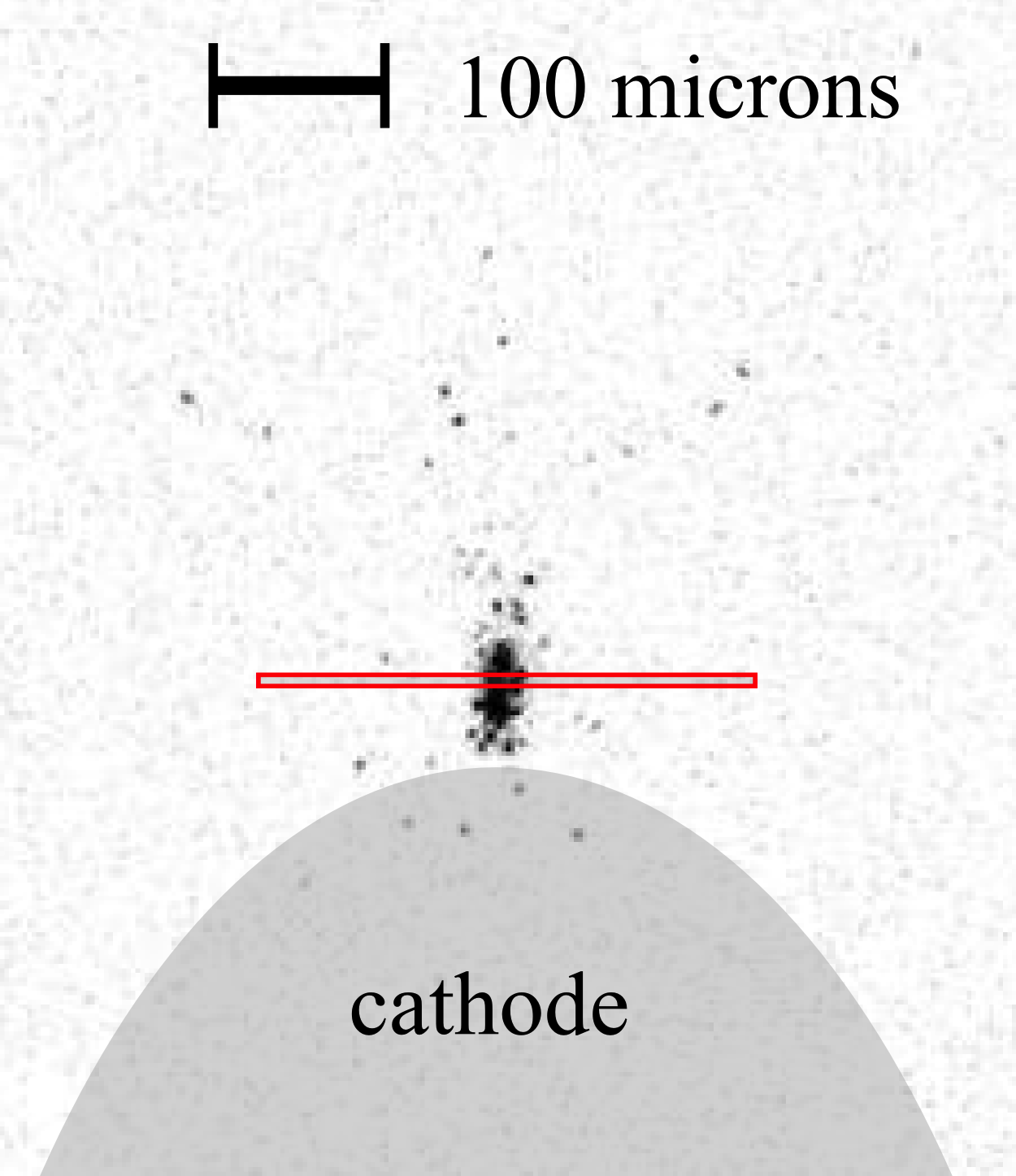}
\end{center}
\caption{Intensified CCD single-shot image of the positive streamer breakdown in the cathode-sheath of the negative corona Trichel pulse at atmospheric pressure air, see also \cite{hoder2012}. {The ten-micrometers wide red slit depicts the area for the temporally resolved FNS and SPS signal detection, as shown in Fig. \ref{exp}.
}
}
\label{iccd}
\end{figure}

The experimental setup for the measurements on Trichel pulse of negative corona discharge streamer was the same as the one used in \cite{hoder2012}. 
It consisted of a grounded cathode with a tip curvature of 190$\,\mu$m and a positive dc voltage (+7.8$\,$kV) connected plate, both made of stainless steel with a gap of 7$\,$mm. This setup resulted in pulses with a frequency of approximately 200$\,$kHz and a current amplitude reaching 4$\,$mA. 
A single shot of the breakdown event is shown in figure \ref{iccd}. 
For the case of streamer measurements in asymmetric barrier discharge (one metallic electrode and the other covered by dielectrics \cite{hoder2010}), the setup was the following: The applied sinusoidal voltage has amplitude of 11$\,$kV$_{p-p}$ (the metal electrode was powered, while the dielectric electrode grounded) and  frequency of 60$\,$kHz. As a dielectrics, an alumina of 96\% purity was used and the discharge gap was 1$\,$mm. 
The spatial resolution of spectroscopic measurements was not worse than 50$\,\mu$m.
In both setups the air flow was of 300$\,$sccm.

The spatio-temporal highly-resolved emission 
was recorded by detection system based on TC-SPC. 
{
For selected spatial coordinate, the TC-SPC method substitutes the real-time emission-measurement of the discharge event by a statistically averaged determination of cross-correlation function between two optical signals, both originating from the same source. These are the so-called `main signal' (spatially and spectrally resolved single photon, the actual detection of the streamer emission) and the `synchronizing signal' (integral light intensity pulse of the microdischarge-event which gives the time-reference). The time between the detections of these signals is measured. Consequently, time histograms of counted photons for all spatial positions of the discharge are accumulated.} 
The TC-SPC detection instrument consists of a time correlated single photon counting module (Becker\&Hickl SPS-150) and two high-sensitive photomultipliers (Hamamatsu PMC-100-4) combined with a monochromator (Acton SpectraPro-500) \cite{hoder2010}. 
The temporal resolution was 12$\,$ps which is the technical division of used TC-SPC memory box. The ICCD image was taken by nanosecond gated camera (DiCam Pro 25 SVGA from PCO Imaging) via a far-field microscope (Questar QM 100BK7).

\begin{figure}[h!]
\begin{center}
\includegraphics[bb = 50 45 750 600,clip = true,width=0.45\columnwidth]{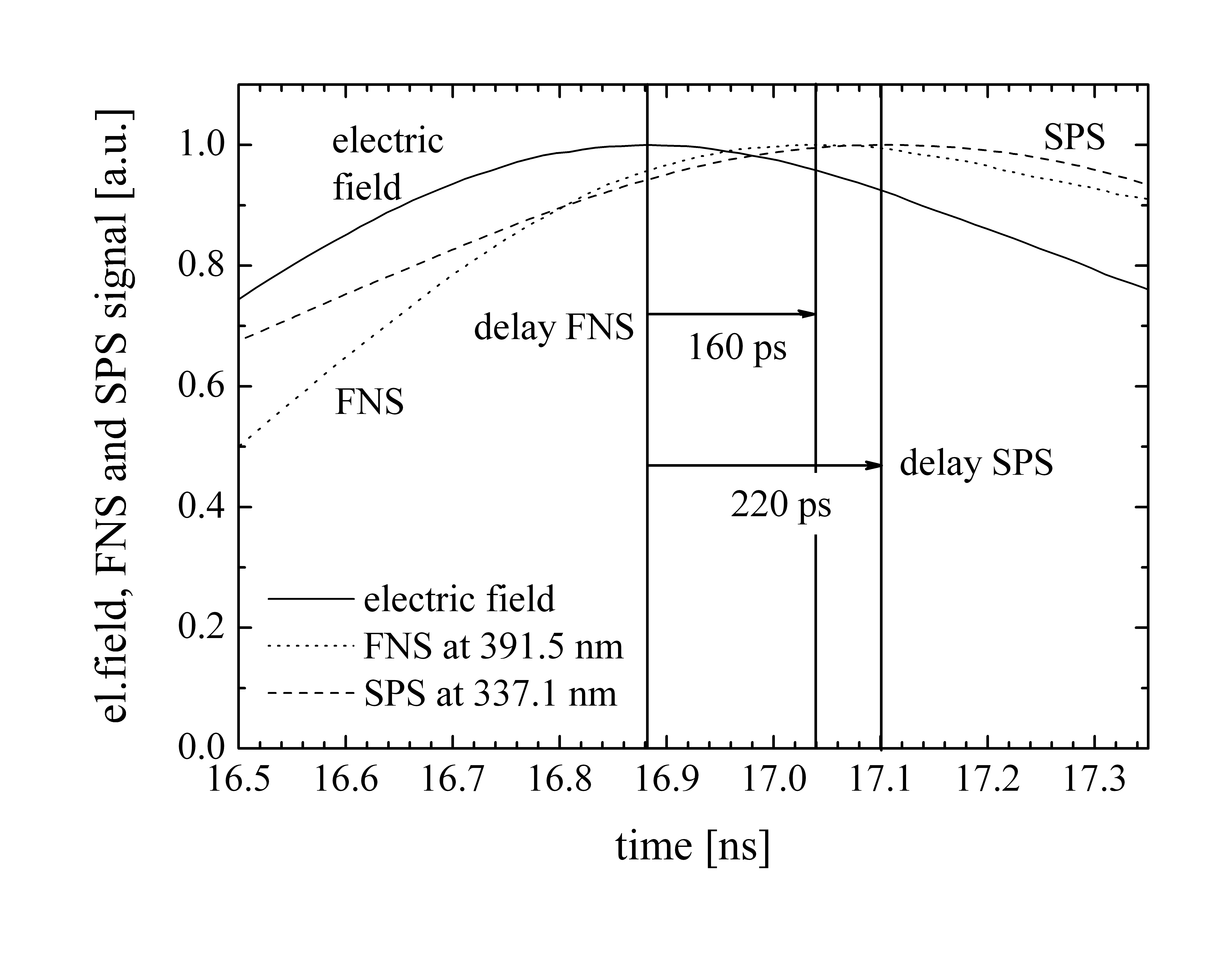}
\end{center}
\caption{Experimentally obtained FNS and SPS signals of positive streamer in its early stage together with determined electric field development. Delays of the FNS and SPS signals maxima to the electric field maximum are denoted. The uncertainty of the obtained delay values is not worse than $\pm 20\,$ps.}
\label{exp}
\end{figure}

In figure \ref{exp}, experimentally obtained FNS and SPS time-resolved emissions of the positive streamer in negative corona Trichel pulse in atmospheric pressure air are shown. They were collected from the red-marked area as shown in figure \ref{iccd}. The kinetic scheme presented in \cite{paris2005,kozlov2001,hoder2012} results in following equation:
\begin{equation}
\label{computation}
\frac{I_{FNS}/\tau_{eff}^{FNS}+\mathrm{d}I_{FNS}/\mathrm{d}t}{I_{SPS}/\tau_{eff}^{SPS}+\mathrm{d}I_{SPS}/\mathrm{d}t}
\cdot
\frac{\tau_{eff}^{FNS}}{\tau_{eff}^{SPS}} 
= R_{\mathrm{FNS/SPS}}(E)
\end{equation}
where the letter $I$ denotes the highly-resolved measured light intensity of the given radiative state in the streamer.
$\tau_{eff}$ is denoting the effective lifetimes and these are given in~\cite{valk} and computed from data in~\cite{valk,dilecce2007,dilecce2010}. 
The dependency $R_{ \mathrm{FNS/SPS}}$ on the electric field $E$ was estimated experimentally from the emission of the non-self-sustaining Townsend discharge with known applied voltage and electrode gap. This was measured in single-table setup for identical parameters and adjustments of the detecting device as for the recording of streamer emission.
Furthermore, spatially resolved spectra were taken with spatial resolution of 10\,$\mu$m and the intensities of the FNS signal were corrected on the overlap with the neighbouring spectral band of SPS (transition 2-5).

The development of the electric field strength was determined and its normalised value is shown in Fig. \ref{exp} as well. 
It is important to note that the primary objective of this work is to investigate mutual delays between
the occurrence  of the electric field maximum  and the optical emissions maxima.
The discussion on the absolute magnitude of the electric field  determined from the optical emissions is
out of scope of the present paper. 
It is apparent from Fig. \ref{exp} that the maxima of the FNS and SPS emissions are delayed differently with respect to the determined electric field maximum, by 160$\,$ps and 220$\,$ps, for FNS and SPS respectively. 
Converting these plots using the propagation velocity of the streamer into the spatial structure of the streamer head one can obtain a picture similar to that shown in \cite{naidis} where it is described as the excitation rate maximum shift. 
Emissions presented in Fig.~\ref{exp} were measured close to the electrically highly-stressed region in needle-cathode vicinity where the velocity of the streamer was approx. $6\cdot10^4\,$ms$^{-1}$. 
The FNS and SPS radially integrated light emissions were collected from identical position through a slit of 10$\,\mu$m width in axial direction.
The experimentally obtained radius of the streamer is 20 $\pm$ 4 $\mu$m which was estimated as an average value from single-shot ICCD images similar to Fig.$\,$\ref{iccd}.
In our case, on its very short path of about 90$\,\mu$m the streamer is in its early stage of development.

\begin{figure}[h!]
\begin{center}
\includegraphics[width=0.45\columnwidth]{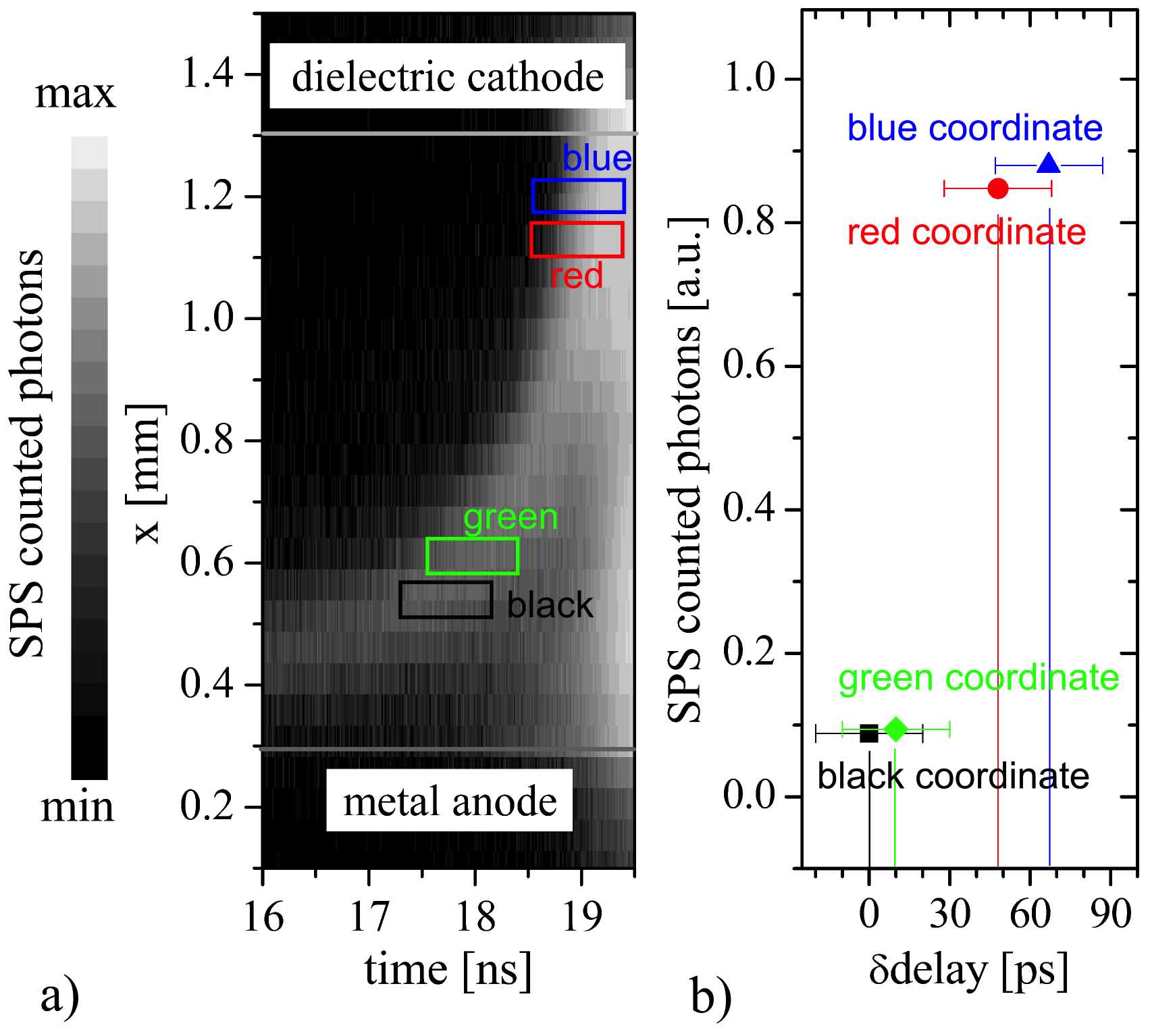}
\end{center}
\caption{Experimentally obtained SPS signal of positive streamer propagating towards dielectric cathode in barrier discharge arrangement (a) with depicted coordinates for the estimation of the delay dilatation. 
The dilatation of the delay $\delta delay$ as well as the increase of the emitted intensity of the SPS signal from the streamer head is shown in part (b).}
\label{dilatation}
\end{figure}

In the barrier discharge setup we observed the same delay phenomena for the FNS and SPS streamer emissions. Moreover, due to the larger electrode gap of 1 mm, we could observe a change in the delay duration as a function of the distance of the streamer head to the cathode. We describe this phenomenon for early-stage streamer emissions as a dilatation of emission delay. 
In Fig.$\,$\ref{dilatation}, the results on the delay dilatation measurements in different
stages of streamer development are shown. In Fig.$\,$\ref{dilatation} a) the spatio-temporal
development of the SPS signal generated by the positive streamer in the
inter-electrode gap of barrier discharge is shown. At the coordinate x =
0.4$\,$mm and t = 16$\,$ns the streamer starts close to the metal anode and
propagates towards the dielectric cathode with increasing velocity. At the
spatio-temporal coordinate ``black" the SPS signal maximum is measured and
its delay value behind the electric field is taken as a reference for the
delays measured in latter stages of the streamer development.  Thus, the delay
dilatation ``black" is set to be equal zero in Fig.$\,$\ref{dilatation} b). The dilatation of the delay
parameter is longer in latter stages of the streamer development.
Spatio-temporallly localised measurements were taken in ``green", ``red" and ``blue"
coordinates and the further change of the delay parameter of the SPS
signal is shown in Fig. \ref{dilatation} b). This delay dilatation is approx. +50 and +70$\,$ps for the ``red" and ``blue" coordinate, respectively. At the same time the increase of the SPS signal intensity maximum for given coordinates is shown. 
To summarize: it is obvious that in the latter stage of the streamer development the 
streamer head SPS emission is more intense with a longer 
dilatation of the delay.

It is important to note that many parameters may influence streamer characteristics and thus the delays. 
For example, as key parameters, we can mention the repetition frequency of the discharge \cite{nijdam2011,hoder2012pop,nijdam2014}, the electrode setup geometry and the air pressure.
Indeed, under transient luminous event conditions in upper atmosphere the emission delays are expected to be observable as well \cite{celestin2010}.
However, for low pressure conditions, the delay should be scaled up towards longer durations in comparison to atmospheric pressure discharges. 
As delays in emission maxima and the dilatation of the delays occur for several streamer discharge conditions in air, in this work, as a first step, we have chosen to study them theoretically with a 2D model case of an air streamer discharge generated in the vicinity of a high voltage spherical electrode immersed into an external homogeneous electric field.

\section{Streamer model}

\begin{figure}[htb]
 \begin{center}
 \includegraphics[width=0.3\columnwidth]{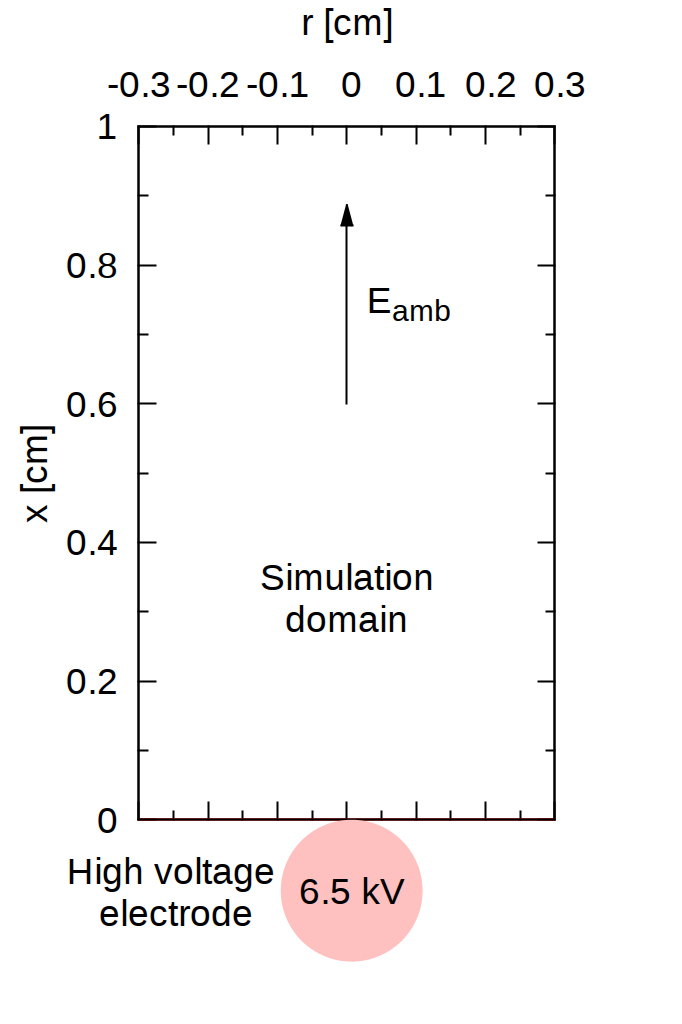}
\caption{Simulation domain: High voltage electrode is a conducting sphere of radius 0.1$\,$cm
         and voltage $6.5\,$kV. Homogeneous ambient electric field $E_{\rm amb}$ of
         $12\,$kVcm$^{-1}$ is established by remote planar electrodes.
         }
 \label{domain}
 \end{center}
\end{figure}

In order to investigate delays in optical emissions 
we have simulated the propagation of positive streamer 
in 2D axi-symmetric geometry in air at atmospheric pressure
using  drift-diffusion equations for electrons, positive and negative ions coupled with
Poisson's equation \cite{babaeva1996}:
\begingroup
\renewcommand*{\arraystretch}{1.5}
\begin{equation}\label{eqn:ne}
\begin{array}{l}
\partial_t n_{\rm e}-\vec{\nabla}{\cdot( n_{\rm e} \mu_{\rm e} {\vec E})}
-\vec{\nabla} \cdot ( D_{\rm e} \vec{\nabla} n_{\rm e} )
= S_{\rm ph}  +S_{\rm e}^{+}-S_{\rm e}^{-}, \\
\partial_t n_{\rm p} = S_{\rm ph}  +S_{\rm p}^{+}-S_{\rm p}^{-}, \\
\partial_t n_{\rm n}  = S_{\rm n}^{+}-S_{\rm n}^{-},
\end{array}
\end{equation} \endgroup
\begin{equation}\label{poisson}
\epsilon_0 \vec{\nabla}^2\phi=
-{q_{\rm e}} (n_{\rm p} - n_{\rm n}- n_{\rm e} ),
\end{equation}
where subscripts `e', `p' and `n' refer to electrons, positive and negative ions, respectively,
$n_i$ is the number density of species $i$, $\phi$ is the electric  potential, $\vec{E} = -\vec{\nabla} \phi$ is the electric field,
$D_{\rm e}$ and $\mu_{\rm e}$ are the  electron diffusion coefficient and the absolute value of electron mobility,
$q_{\rm e}$ is the absolute value of electron charge, and  $\epsilon_0$ is permittivity of free space.
The $S^{+}_i$ and $S^{-}_i$ terms stand for the rates of production and loss of charged particles.
The $S_{\rm ph}$ term is the rate of electron-ion pair production due to photoionization in a gas volume.
Reaction rates and transport coefficients for air are assumed to be functions of
the local reduced electric field $E/N$ , where $E=|\vec{E}|$ is the electric
field magnitude and $N = 2.688\times10^{25}\,$m$^{-3}$ is the air neutral
density. 
The transport and source parameters
are taken from \cite{morrow1997}. The photoionization
is taken into account through the 3-Group SP$_3$ method
derived by \cite{bourdon2007} and \cite{liu2007}.
Note that on timescales of interest for this work,
ions are considered motionless.
Positive streamer is initiated by placing a  Gaussian plasma cloud 
with a peak density of 
$10^{18}\,$m$^{-3}$ and a characteristic length scale $10^{-4}\,$m
in a high-field region in the vicinity of a conducting sphere of radius $0.1\,$cm with an 
applied potential of $6.5\,$kV, see Figure$\,$\ref{domain}.
The sphere is immersed in a homogeneous electric field $E_{amb}$ that ranges between $8$ and $18\,$kVcm$^{-1}$.
Similar  configuration was considered for derivation
of a correction factor stemming from geometrical shape of luminous streamer heads
for sprite conditions in \cite{celestin2010}  and for streamers
at ground pressure air in \cite{bonaventura2011}.
For the sake of brevity we just point out
that, in this work, we have used a $1.0\times0.3\,$cm$^2$ (i.e., length $\times$ radius)
computational domain discretised on a fixed rectilinear
grid with a mesh size of 6.2$\,\mu$m.
More details about the model can be found in \cite{bonaventura2011}.
Detailed justification for similar fluid model of streamer propagation, including
disussion on limits of its validity,  is presented in \cite{LuqueJCP2012}. Moreover,
direct comparision of results of PIC-MCC model \cite{ChanrionJCP2008} for streamer
propagation and results of fluid model \cite{liu2004} also shows excellent agreement \cite{ChanrionJCP2008}.

\begin{figure}[htb]
 \begin{center}
 \includegraphics[width=1\columnwidth]{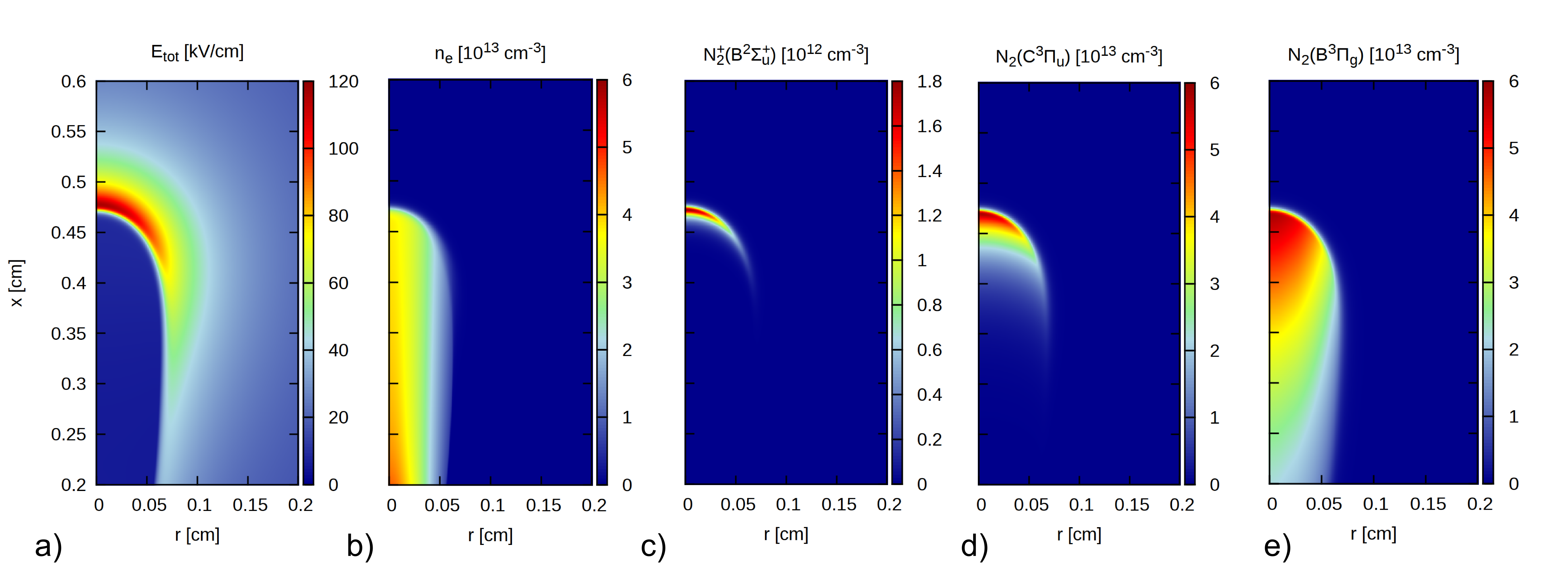}
\caption{Cross sectional views of electric field (a), electron density (b), FNS (c), SPS (d) and FPS (e) for a positive streamer at time $t = 10.0\,$ns when it approaches the middle of the simulation domain, see Fig. \ref{domain}.}
 \label{alltogether}
 \end{center}
\end{figure}

To calculate the optical emissions of the SPS, FNS and also FPS (first positive system) band systems, 
we use a model similar to the
one given in \cite{pasko1997}.
The population of the excited species N$_2$(C$^3\Pi_u$), N$_2^+$(B$^2\Sigma_u^+$) and N$_2$(B$^3\Pi_g$) is governed by:
\begin{equation}
{{\partial n_k \over \partial t}} = -\frac{n_k}{\tau_k} + \nu_k n_{\rm e},
\label{excited}
\end{equation}
where $n_k$ [cm$^{-3}$] is the population   of excited state $k$,
and $\nu_k$ is  the frequency of creation of excited state $k$  by electron impact,
$\tau_k = [A_k + \alpha^k_{\rm N_2} N^{\phantom k}_{\rm N_2} +
                       \alpha^k_{\rm O_2} N^{\phantom k}_{\rm O_2} ]^{-1}$
is the total lifetime of $k$-state,
$\alpha^k_{\rm X}$ is a quenching rate of $k$ due to collisions with molecule of type X of
 density $N_{\rm X}$ and $A_k$
[s$^{-1}$] is the Einstein coefficient. 
The quenching rates and Einstein coefficient sets from \cite{liu2004} are used throughout this work.
The equation (\ref{excited}) for excited states is solved with the streamer equations (\ref{eqn:ne})--(\ref{poisson}).
This gives a full time-dependent solution of optical emissions in the modelling of the streamer processes, see Fig.$\,$\ref{alltogether}.

Intensity  of light emission $I_k$ of a state $k$
is proportional to radiative dexcitation rate $A_k$ (s$^{-1}$):
\begin{equation}
I_k = A_k n_k.
\end{equation}
Intensity of light emitted from a discharge is usually line-of-sight (LoS) integrated, 
then the LoS optical emission  intensity  of state  $k$ is given by
\begin{equation}
\Psi_k=10^{-6}\int_{\cal L} I_k\,\d l, \label{LoSIntensity}
\end{equation}
where $I_k$ is in cm$^{-3}$s$^{-1}$, length $l$ of the optical path ${\cal L }$ is in cm
and $\Psi_k$ is in Rayleighs.
The effect of radiative transfer between  the source of the emission
and the detector  is not taken into account.

\section{Results and discussion}

\subsection{Causes of the emission delays}

\begin{figure}[h!]
\begin{center}
\includegraphics[width=0.45\columnwidth]{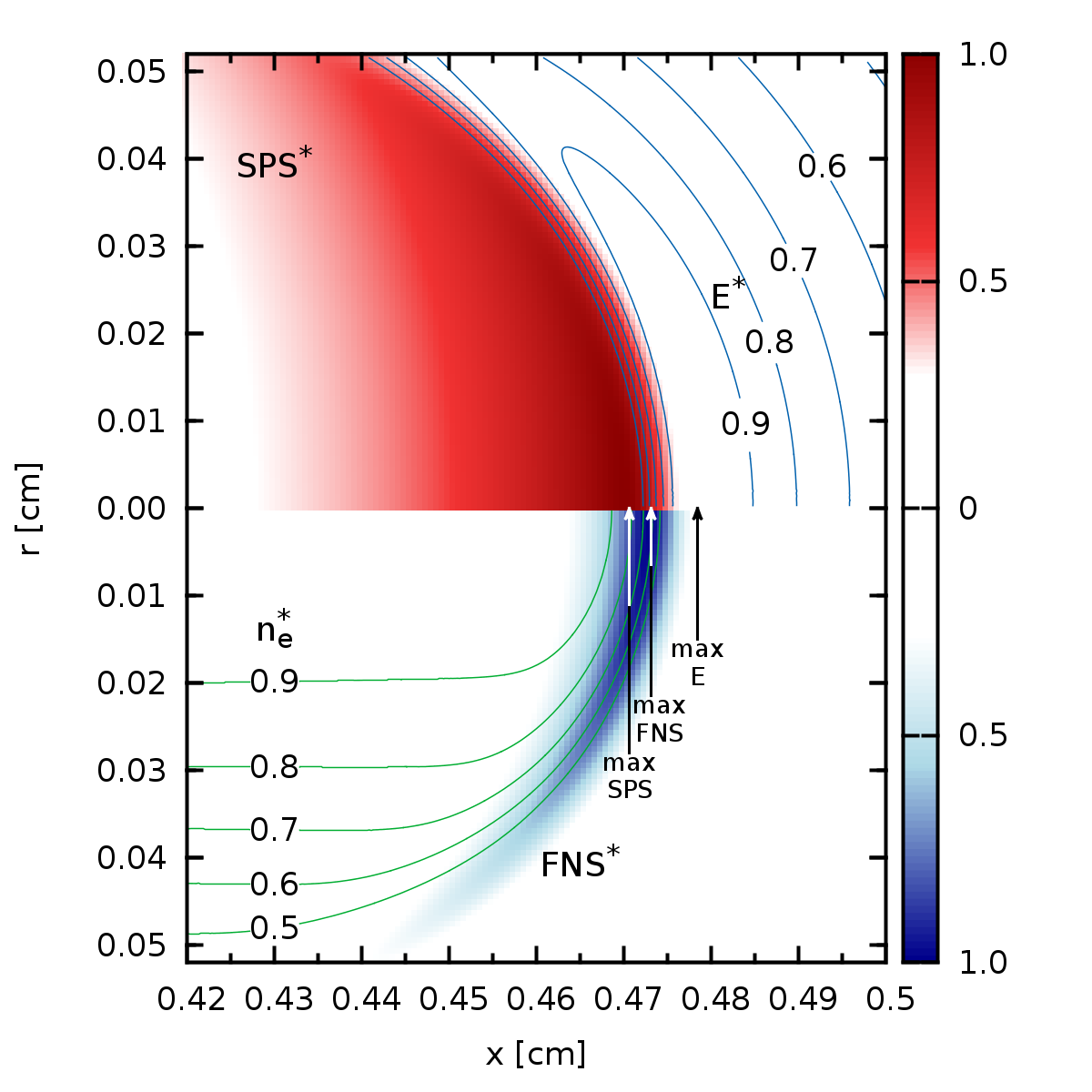}
\end{center}
\caption{Detailed view on the positive streamer head structure obtained from 2D axi-symmetric simulation. In upper part, electric field isolines and emission profile of SPS (red) are shown. In the part below, electron density isolines and emission profile of FNS (blue) are presented. The star-signs denote a normalised value. Note, that while the results of measurements on the streamer discharge are typically represented in the time scale of measured temporal development, i.e. from left to the right as in Fig.$\,$\ref{exp} or in \cite{bonaventura2011,kozlov2001,shcherbakov2007,hoder2012}, the results of numerical modelling are presented in opposite direction \cite{naidis,celestin2010}, which is also the case further in this article.
}
\label{head}
\end{figure}

The two-dimensional plots of the development of the electric field, FNS and SPS emission and electron density were evaluated from the simulations (see Fig. \ref{alltogether}). 
For these conditions, 10$\,$ns after ignition, the positive streamer is propagating 
in homogeneous electric field of 12$\,$kVcm$^{-1}$ with velocity of about $5\cdot10^5$ ms$^{-1}$. 
In two dimensions (axial and radial), streamer head distributions of SPS and FNS emission intensities together with electric field and electron density in the $x - r$ plane are shown in Fig.$\,$\ref{head}. 
The position of the signal and parameter maxima in 2D streamer head profile are denoted. 
From Fig.$\,$\ref{head} one can see that as the streamer head, i.e. the maximum of the electric field, passes the spatial coordinate, the FNS maximum follows with the delay of 100$\,$ps (83 $\mu$m at given velocity) and after next 50$\,$ps (55 $\mu$m) the maximum of SPS emission follows. 
Note that these are axial values obtained from radially resolved simulations thus not directly comparable to the experimentally obtained (radially unresolved) ones.

\begin{figure}[h!]
\begin{center}
\includegraphics[width=0.8\columnwidth]{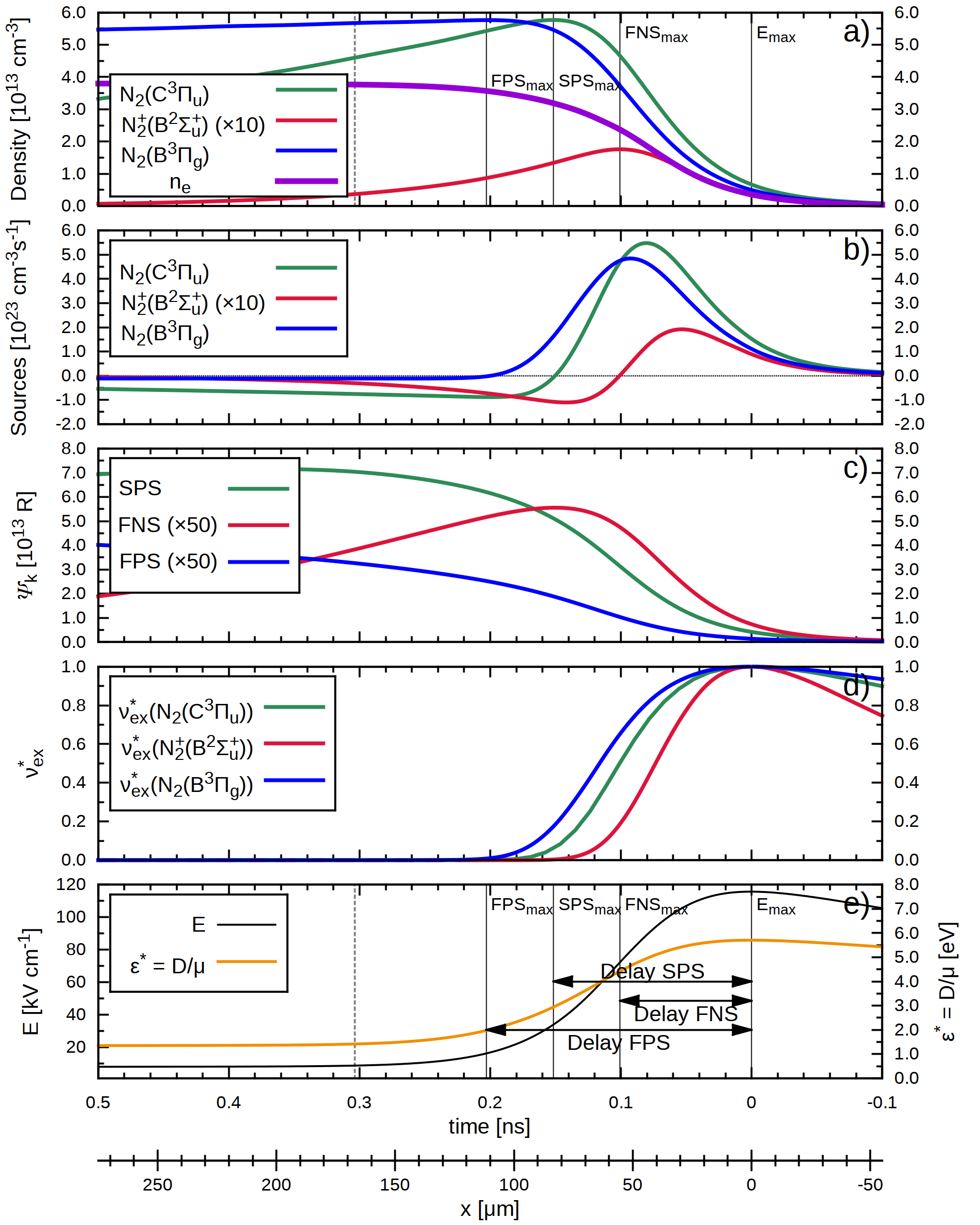}
\end{center}
\caption{Time (space) evolution of the axial densities of N$_2$(C$^3\Pi_u$), N$_2^+$(B$^2\Sigma_u^+$) and N$_2$(B$^3\Pi_g$) states and electrons $n_{\mathrm{e}}$ are shown in panel (a).
Source terms ${{\partial n_k / \partial t}}$, $\Psi_{k}$ (line-of-sight) intensities and normalized excitation rates  $\nu^*_{\mathrm{ex}}$ for all three excited states are given in panel (b), (c) and (d), respectively. In the panel (e), development of $x$-component of the electric field E and characteristic electron energy $\varepsilon$* = D/$\mu$ is shown. 
Full vertical lines indicate positions of peak axial values of the electric field (E$_{\rm{max}}$) and emission intensities (SPS$_{\rm{max}}$, FNS$_{\rm{max}}$ and FPS$_{\rm{max}}$).
The delays marked by horizontal arrows have values of 150$\,$ps (83$\,\mu$m), 100$\,$ps (55$\,\mu$m) and 200$\,$ps (112$\,\mu$m) for SPS, FNS and FPS signal maxima, respectively. 
Vertical dashed line 
marks a position where a radially resolved spectrum is analysed in Section 4.3. 
The spatial and time coordinates are linked via streamer velocity of 5$\cdot 10^5\,$ms$^{-1}$.
}\label{sources}
\end{figure}

In order to understand these delays the population evolutions of the radiative states, 
their source terms (see equation \ref{excited}) and other related characteristics are 
visualised in figure \ref{sources}. There, the development 
of electron density and electric field is shown in panels a) and e). 
Optical emissions rise sharply reaching peak values after the 
E$_\mathrm{max}$ together with an increase of $n_{\rm{e}}$, 
because the population of corresponding excited states is determined by the product of $n_{\rm{e}}$ and excitation rate.
The characteristic energy $\varepsilon^*$ of electrons (see panel e)) evolves quickly achieving its maximum of approx. 5$\,$eV nearly synchronously with the peak of electric field E$_\mathrm{max}$ and decreasing towards streamer channel stationary value of approx. 1.3$\,$eV within 200$\,$ps. Such a drop implies significant variations of the EEDF characterized by a much faster disappearance of high energy (tail) electrons and, in general, determines characteristic timescale of the population dynamics of electronically excited states.
The position of FNS and SPS maxima is governed by the balance established between gain and loss terms. 
Behind the streamer head (i.e. the coordinate E$_\mathrm{max}$) where both $E$ and $\varepsilon^*$ decrease, the creation frequency for FNS $\nu_{FNS}$ 
decreases faster (insufficient electrical field for accelerating electrons over the FNS excitation threshold of $18.8\,$eV) than $\nu_{SPS}$ (SPS threshold 11$\,$eV), see panel d).
The difference between two excitation thresholds is imprinted via different rates for FNS and SPS excitations (panel d)).
Consequently, the SPS remains sufficiently excited even in low field region where the FNS source term vanishes. 
Moreover, FNS has much shorter effective lifetime than SPS (effective lifetime of SPS is 0.61$\,$ns and 0.12$\,$ns for FNS under given conditions \cite{liu2004} or even shorter for FNS as reported in \cite{dilecce2010}).
This scenario results in sharper FNS emission peak which appears closer to the electric field maximum than more distant and broader peak of SPS.

For comparison, in Fig.$\,$\ref{sources}, the development of N$_2$(B$^3\Pi_g$) axial population (FPS emission) is presented as well. 
Comparing FPS, SPS and FNS maxima delays, the delay of FPS emission maximum with respect to the peak of electric field is the longest one (about 200$\,$ps). 
This is because the threshold for electron impact excitation of the N$_2$(B$^3\Pi_g$) state (FPS emission) is significantly lower (7.4$\,$eV) compared with excitation threshold of N$_2$(C$^3\Pi_u$) and N$_2^+$(B$^2\Sigma_u^+$). 
Moreover, the lower excitation threshold of the N$_2$(B$^3\Pi_g$) is responsible for much higher population density than in the case of N$_2^+$(B$^2\Sigma_u^+$). 
The form of the FPS emission correlates with the electron density distribution more closer than any other presented spectral system emission, compare the sub-figures b), c), d) and e) in Fig.$\,$\ref{alltogether}.

To summarise, sub-nanosecond delays of peak intensities of the SPS, FNS and FPS systems with respect to E$_\mathrm{max}$ 
signal maxima behind the streamer head are caused by different excitation threshold energies of excited states, different radiative and collisional quenching, 
as well as by fast decrease of characteristic energy of electrons $\varepsilon$* and slowly increasing electron density $n_{\rm{e}}$ behind  
the streamer head.

\subsection{Streamer geometry and its fingerprint in spectral signatures}

\begin{figure}[h!]
\begin{center}
 \includegraphics[width=0.49\columnwidth]{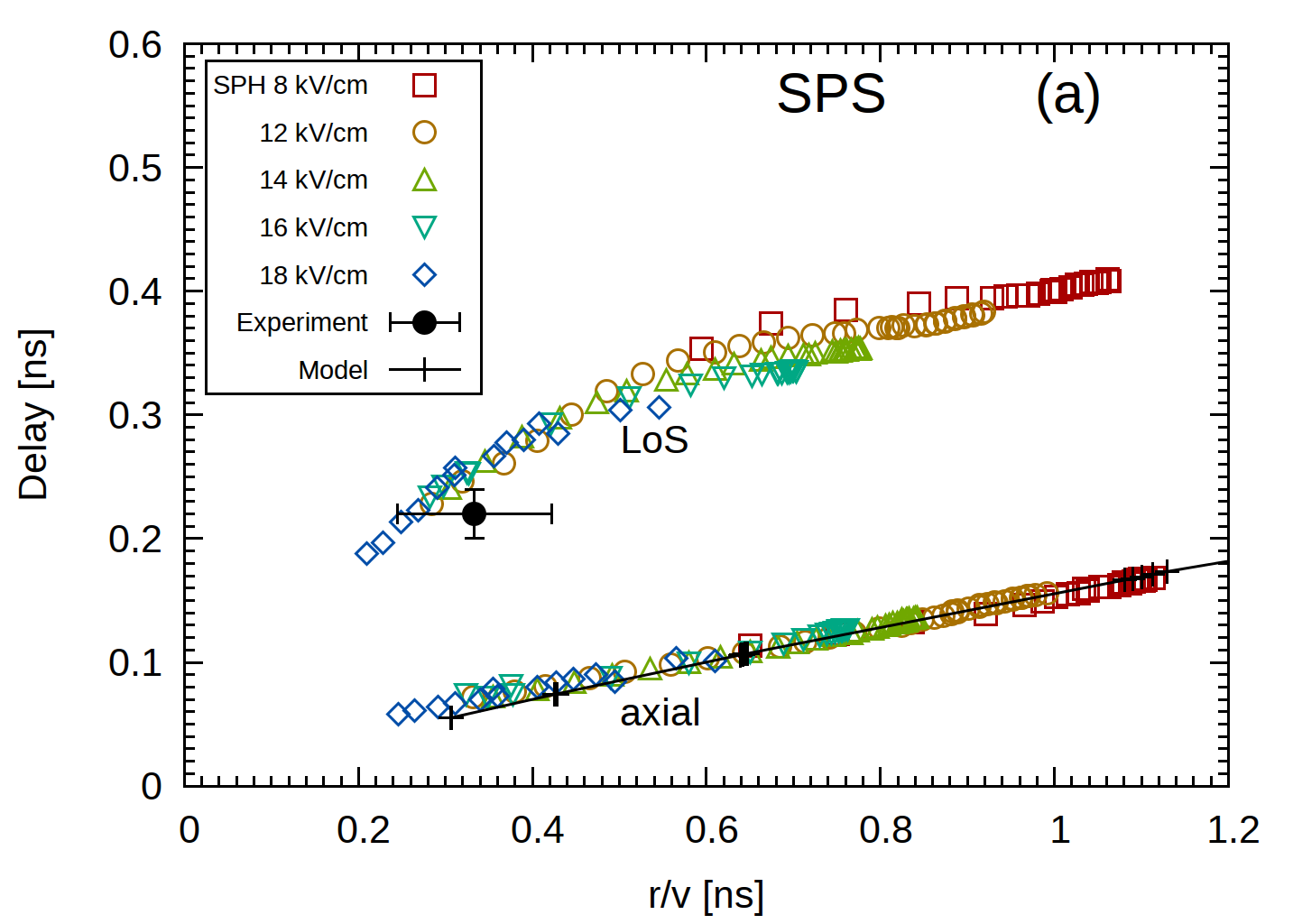}
 \includegraphics[width=0.49\columnwidth]{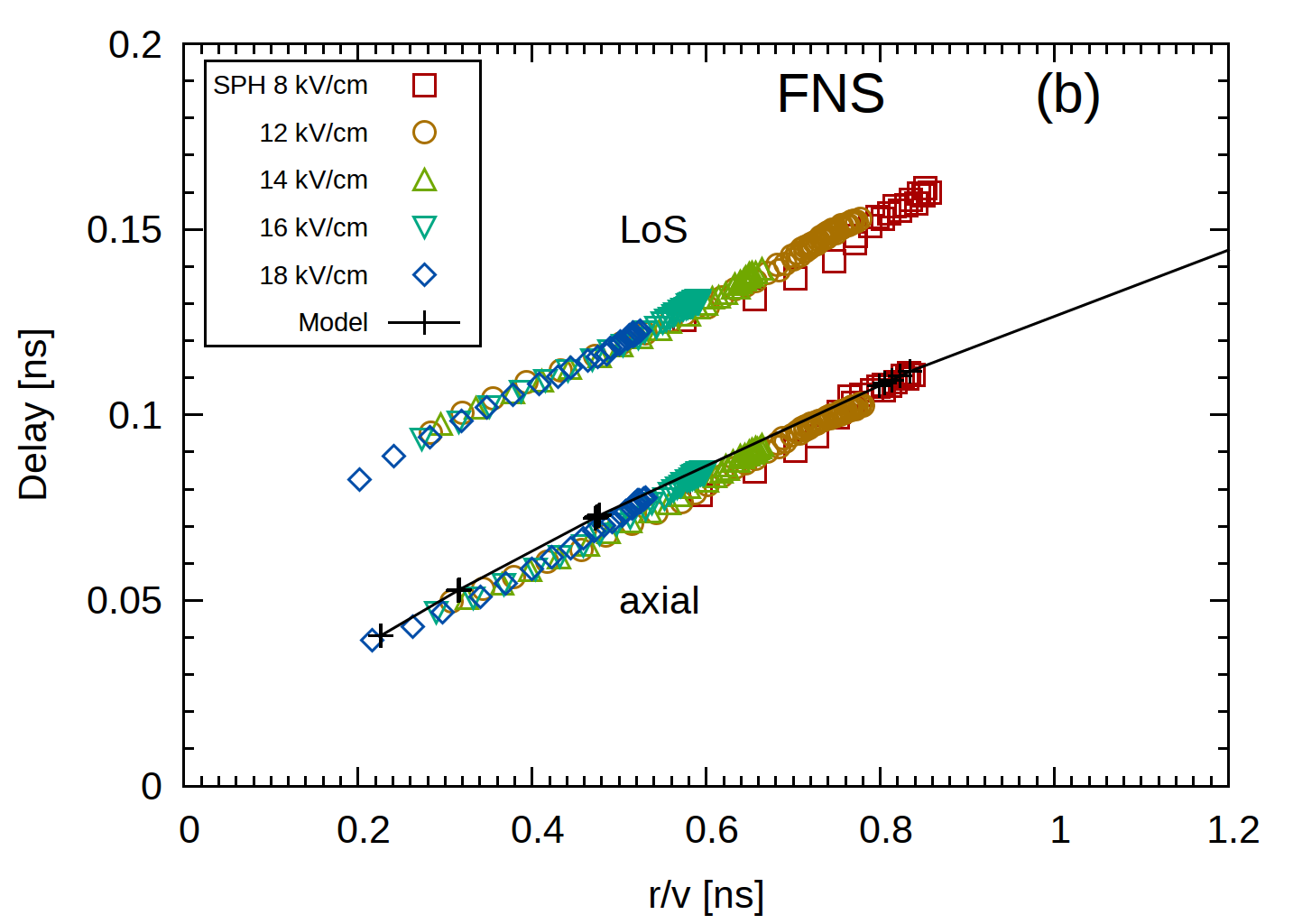}
\end{center}
\caption{Delays of optical emission peaks for SPS (a) and FNS (b) for positive streamer as a function of
streamer radius $r$ and streamer velocity $v$  ratio for different values of the homogeneous ambient electric field (see Fig.$\,$\ref{domain}). Streamer radius is determined as: 1)
the (radiation) radius where density of excited states is 1/2 the value on the axes for `axial' group of points;
2) the radius where the line-of-sight (LoS) radially integrated  intensity is 1/2 of the maximum value
for `LoS' group of points. Points are results of 2D axi-symetric simulation and
lines are modelled by eq.(\ref{tlag}), see further in the text.}
\label{delayrv}
\end{figure}

The dependence of the computed delays (SPS and FNS) on the ratio of the streamer radius and velocity $r/v$ at given moments of its development are shown for different values of the homogeneous ambient electric field (Fig.$\,$\ref{delayrv}). 
Individual points in the figure represent individual instants of the
streamer propagation. Generally all points for each set are ordered in
time when seen from left to right. 
In other words, points of given colour lying more on the left correspond to the earlier phases of the streamer development.
Clearly, a linear trend can be observed for the set of `axial' values (local emission from the axis of the propagating streamer).

Considering first the delays of peak densities of excited states on the axis of the discharge, 
we see that no matter what model or external field, all points fit a single line (`axial'). 
Clearly the ratio $r/v$ is
a key characteristic of a streamer for discussing the excitation delay.
Obviously, both the delay and $r/v$ are linked characteristics of the
streamer head geometry and thus connected to more fundamental features of the streamer structure. 
On the other head, the so called line-of-sight (LoS, i.e., intensity integrated through the whole streamer diameter at selected $x,r$-coordinate 
) delays dependence on $r/v$ is a bit more complex. We see convergence to 
a single path for the LoS points that correspond to instants of early
streamer development (points more on the left for each set). For more developed streamers (points more on the right of each set) a dispersion of the delays obtained for various simulation conditions is observed.

Even though we do not intend to simulate in this work the particular case of the Trichel pulse streamers presented in Section 2, from Fig.$\,$\ref{delayrv} one can see that the experimentally observed delay value (at least for SPS, as we have no experimental value for the FNS emission radius) is not far from the simulated one when we consider the both streamers, the measured and the simulated one, in their early development stage. 
Note that in the experiment the emission 
is projected and integrated over the whole streamer diameter (direct Abel transformation) at selected $x$-coordinate, while for LoS emission the emission is integrated through 
the projection along single line-of-sight which is passing streamer axis ($r$ = 0). 
To radially resolve the 40$\,\mu$m thin streamer by maintaining the high temporal resolution simultaneously is over the possibilities for given setup.

From the results presented in Fig.$\,$\ref{delayrv} one can also conclude that the later is the stage of streamer development the longer are the delays of the peak emissions. 
SPS recordings for the positive streamer propagating along the 1$\,$mm long gap in dielectric barrier discharge \cite{hoder2010} was analysed in previous section (see Fig.$\,$\ref{dilatation}). 
Indeed, a delay dilatation was observed. After 500$\,\mu$m of the streamer propagation the dilatation was +50$\,$ps and 100$\,\mu$m in front of the cathode reaching already +70$\,$ps for the SPS delay. 
The obtained result is consistent with the simulated delay dilatations.
This is due to the spatial scale of the streamer head which expands in time both in the experiment as well as in presented models.

Strongly nonlinear features of streamers are well known and thoughtfully discussed in many works, see,
e.g., in \cite{nonlinear,luque2008}. Nevertheless
in this work, based on the results of 2D axisymmetric simulations,
the dependence of the delays on the streamer radius to velocity ratio $r/v$
is found to be linear for the emissions on the streamer axis.
In order to explain this surprising linearity, in the following 
we analyze  the positive streamer in the framework of a 1D analytical approximation.
This model is capable to mimic basic features of the
streamer propagation \cite{kulikovsky1998,kulikovsky1998ieee,naidis} and through its obvious simplicity to
explain  in very comprehensive manner  the meaning of the observed linearity.
Dependence of $E$ on $z$ along the axis in the vicinity of the streamer 
head may be reasonably  approximated by an expression
\begin{equation}
\label{EzField}
 E(z) =  \cases{ \Es(1+z/\lf)^{-1} &for  $z > 0$, \cr
                   \Es(1+2z/\lf)     &for  $ -\lf < z < 0$,}
\end{equation}
where $\Es$ is the peak electric field in the streamer head at position  $z=0$,
and  $\lf$ is the axial width of the streamer high field region \cite{kulikovsky1998}.
The simplest equation describing a streamer propagation is based on a fluid approximation, neglecting
photoionization and diffusion and has a form:
\begin{equation}
\label{1DTransport}
\frac{\partial \ne }{\partial t} + \nabla\cdot(\ne \vec\we)=\alphaeff\we,
\end{equation}
where $\vec \we$ is the electron drift velocity, $\alphaeff$ is the effective 
ionization coefficient, both functions of $E$. 
Assuming that the streamer is propagating with a constant velocity $\vs$, 
then, following \cite{naidis}, we can integrate 
(\ref{1DTransport}) in the framework of a 1D approximation by using (\ref{EzField}) for $z > 0$ to get
\begin{equation}
\label{ne:infornt}
\nes(\vs+\wes)=\neb(\vs+\web)\exp\left(\lf\Es\int_\Eb^\Es \frac{\alphaeff \we}{\vs+\we}\frac{\d E}{E^2}\right),
\end{equation}
where $\Eb = 30\,$kVcm$^{-1}$ is the magnitude of the electric field on the edge of the ionization region, $\nes$ and $\neb$
are electron densities at the position of the peak electric field ($z=0$) and on the edge of the ionization region
where $E=\Eb$, respectively. Similarly $\wes$ and $\web$ denote electron drift velocities where the electric
field value equals $\Es$ and $\Eb$.
Similarly by using (\ref{EzField}) for $ z < 0$  one gets electron density on the axis of streamer behind the electric peak:
\begin{equation}
\label{ne:behind}
\nec(\vs+\wec)=\nes(\vs+\wes)\exp\left(\frac{\lf}{2\Es}\int_\Ec^\Es \frac{\alphaeff \we\, \d E}{\vs+\we}\right),
\end{equation}
where $\nec$ is electon density behind the electric field peak where the magnitude of the
field diminished to value $\Ec$ and $\wec$ is corresponding electron drift velocity.
Equation (\ref{ne:infornt}) allows to obtain a condition that 
relates the axial streamer width $\lf$ to the streamer velocity 
$\vs$ through the streamer peak field $\Es$:
\begin{equation}
\label{VelocityWidthRelation}
\lf\Es\int_\Eb^\Es \frac{\alphaeff \we}{\vs+\we}\frac{\d E}{E^2} = \ln\left(\frac{\nes}{\neb}\right) + \ln\left(\frac{\vs+\wes}{\vs+\web}\right),
\end{equation}
which allows one to obtain for a given $\Es$ and $\lf$ coherent values of streamer velocity $\vs$,  see \cite{naidis}.

Observed linear dependence of delays of maximum intensity of SPS and FNS emission 
on the ratio $\rs/\vs$ 
can be explained on the basis of equation (\ref{EzField}). Considering 
the dependence $E(z)=\Es(1+2z/\lf)$, 
an observer at a fixed point would find that when the streamer is passing 
its location with velocity $\vs$, then  after the peak, the electric field decays linearly with time 
\begin{equation}
\label{etime}
E(t)=\Es(1-2\vs t/\lf).
\end{equation}
Note also that most of  the excitation of radiating states, in fact, takes place
after the passage of the peak electric field, 
when the electric field starts to diminish. Therefore after some time,
the lost of excited states  overcomes their production.
The maximum of the emission occurs when lost and gain terms in equation (\ref{excited}) are 
equal, i.e. when $n_k/\tau_k=\ne\nu_k$.
The time lag $\tlag$ between the maximum of the electric 
field and the maximum of emission defines the delay and may be estimated by simply
recasting  the equation (\ref{etime}) for time:
\begin{equation}
\label{tlag}
\tlag = \left(1-\frac{E(\tlag)}{\Es}\right)\frac{\lf}{2\vs}.
\end{equation}
Taking into account coherent combination of streamer parameters $\{\Es,\lf,\vs\}$
obtained from condition (\ref{VelocityWidthRelation}) with electron density variation as
given by (\ref{ne:behind}), then (\ref{excited}) can be integrated 
to find time $\tlag$ (or electric field $E(\tlag)$ in (\ref{tlag})) 
when loss and gain terms of (\ref{ne:behind}) equal. 
This model was evaluated
for $\alphaeff$  and $\we$ from \cite{morrow1997} and for $\ln(\nes/\neb)=8$ (see \cite{naidis}) 
for $\Es$ in the range between 100 and 200$\,$kVcm$^{-1}$ with step of 20 kVcm$^{-1}$ and $\lf$ in between 0.05 to 0.125$\,$cm with a step of 0.015$\,$cm.
Note that according to \cite{naidis,kulikovsky1998,babaeva1996,pancheshnyi} ratio of radiation radius $r$
of the streamer to the width  $\lf$ ranges between $\xi=1.5$--$2.5$.
Time delays resulting  from the model (\ref{tlag})
for $\xi=1.7$ in case of FNS and $\xi=2.3$ in case of SPS (solid lines with points) 
together with data obtained from 2D axi-symmetric
simulations (coloured symbols) are presented in Figure \ref{delayrv}.
Finally one can observe that, despite the fact that
the slope $(1-E(\tlag)/\Es)$ in equation (12) generally depends
on a set of $\{\Es,\lf,\vs\}$ values, the overall dependence of $\tlag$
on the radiation radius to velocity ratio is not far from linear.

\subsection{Spectrometric representation of the streamer head structure}

Concentrations of N$_2$(C$^3\Pi_u)$, N$_2^+$(B$^2\Sigma_u^{+})$ and N$_2$(B$^3\Pi_g)$ species (i.e. SPS, FNS and FPS respectively, see figure \ref{alltogether}) 
calculated at any spatial coordinate ($x, r$) in the $x-r$ plane can be represented by corresponding synthetic emission spectrum $i(\lambda, x, r)$ calculated assuming fixed spectral resolution (given by instrumental function of spectrometer) and certain characteristics of emitting states (rotational temperatures and vibrational distributions). Integration of $i(\lambda, x, r)$ synthetic spectra along $r$ (assuming cylindrical symmetry) and/or $x$ coordinates allows evaluating instrumental effects associated with spatial, temporal and spectral resolution limits occurring in real experiments. For example, streamer emission is usually detected through projected luminosity of the streamer filament determined by the unknown radial distributions of various radiating species. Integrating $i(\lambda, x, r)$ along the plane perpendicular to the direction of the streamer propagation ($x$ = const.) therefore simulates radially integrated spectra $I(\lambda, x)$ which are usually used to evaluate streamer parameters. Integrating $I(\lambda, x)$ along x-coordinate then allows accounting for limited temporal resolution of real ICCD or PMT detectors. 

\begin{figure}[h!]
\begin{center}
\includegraphics[bb = 40 40 760 350,clip = true,width=1\columnwidth]{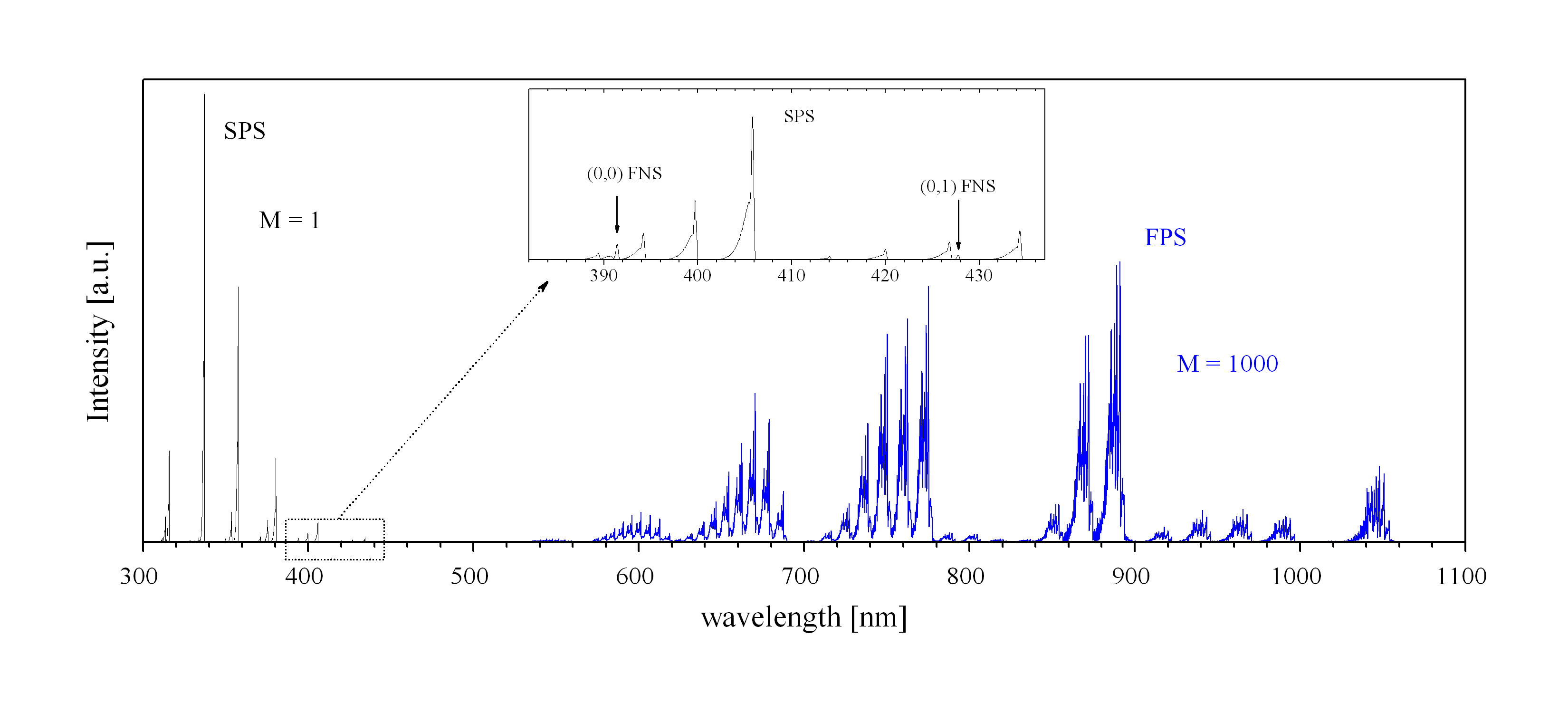}
\end{center}
\caption{Synthetic streamer head emission spectrum integrated over the whole $x - r$ plane displayed  in Fig. \ref{head}. Calculated assuming T$_\mathrm{rot}$ = 300$\,$K and using triangular instrumental function (FWHM = 0.2$\,$nm). The value $M$ denotes the multiplication factor.}
\label{head-spec}
\end{figure}

We applied an approach detailed in \cite{simek2002,simek1995} to construct synthetic SPS, FNS and FPS emission spectra occurring in the 300--1100$\,$nm spectral range by fixing rotational temperature of 300$\,$K for all emitting states and using line-shape defined by triangular instrumental function (spectral resolution of 0.2$\,$nm), for axially and spectrally integrated signal see Fig. \ref{head-spec}.
Because the code which was used to simulate populations of excited electronic states of N$_2$ and N$_2^+$ species does not include vibrational kinetics, i.e. populations obtained represent sum over all vibrational levels of a given state, corresponding data (such as shown in Fig.\ref{head}) can be therefore represented by emission spectra only after assuming certain vibrational distribution for each electronic state.
In the case of the FNS we assumed that electron-impact ionization populates exclusively v = 0 vibrational level of the N$_2^+$(B$^2\Sigma_u^{+})$ state, whereas the N$_2$(C) and N$_2$(B$^3\Pi)$ state populations were distributed among   v = 0--4 (1 : 0.18 : 0.06 : 0.015 : 0.002) and v = 0--12
(0.64 : 1.00 : 0.98 : 0.47 : 0.39 : 0.26 : 0.12 : 0.073 : 0.041 : 0.022 : 0.011 : 0.0055 : 0.0028) vibrational levels, respectively. 
Final $i(\lambda, x, r)$ spectra were constructed 
by blending SPS, FPS and FNS systems according to calculated local $(x, r)$ populations of individual  N$_2$(C$^3\Pi,v)$, N$_2$(B$^3\Pi,v)$ and N$_2^+$(B$^2\Sigma,v)$ vibrational levels, respectively.

\begin{figure}[h!]
\begin{center}
a)~
\includegraphics[bb = 40 60 750 350,clip = true,width=1\columnwidth]{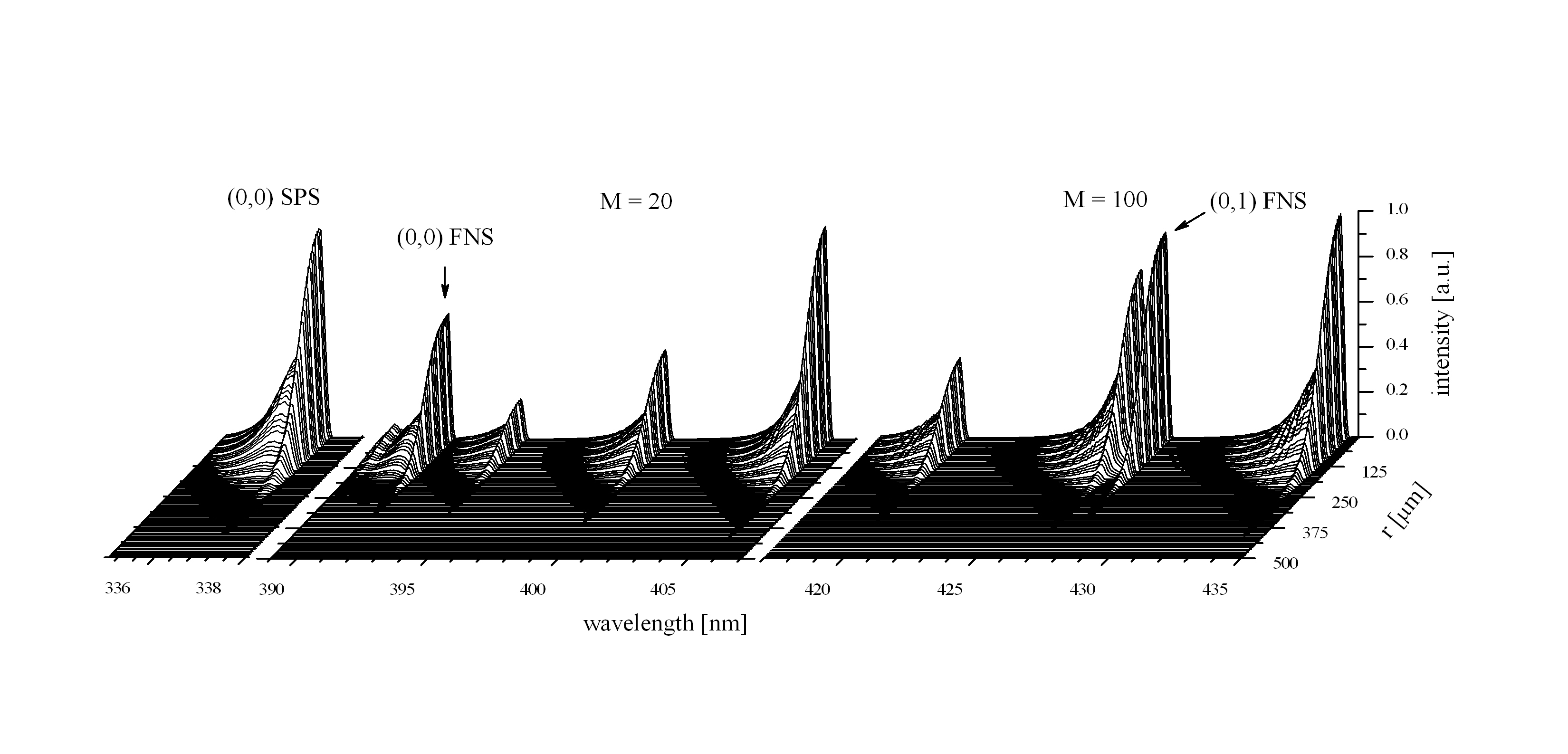}
b)~
\includegraphics[bb = 90 100 750 350,clip = true,width=1\columnwidth]{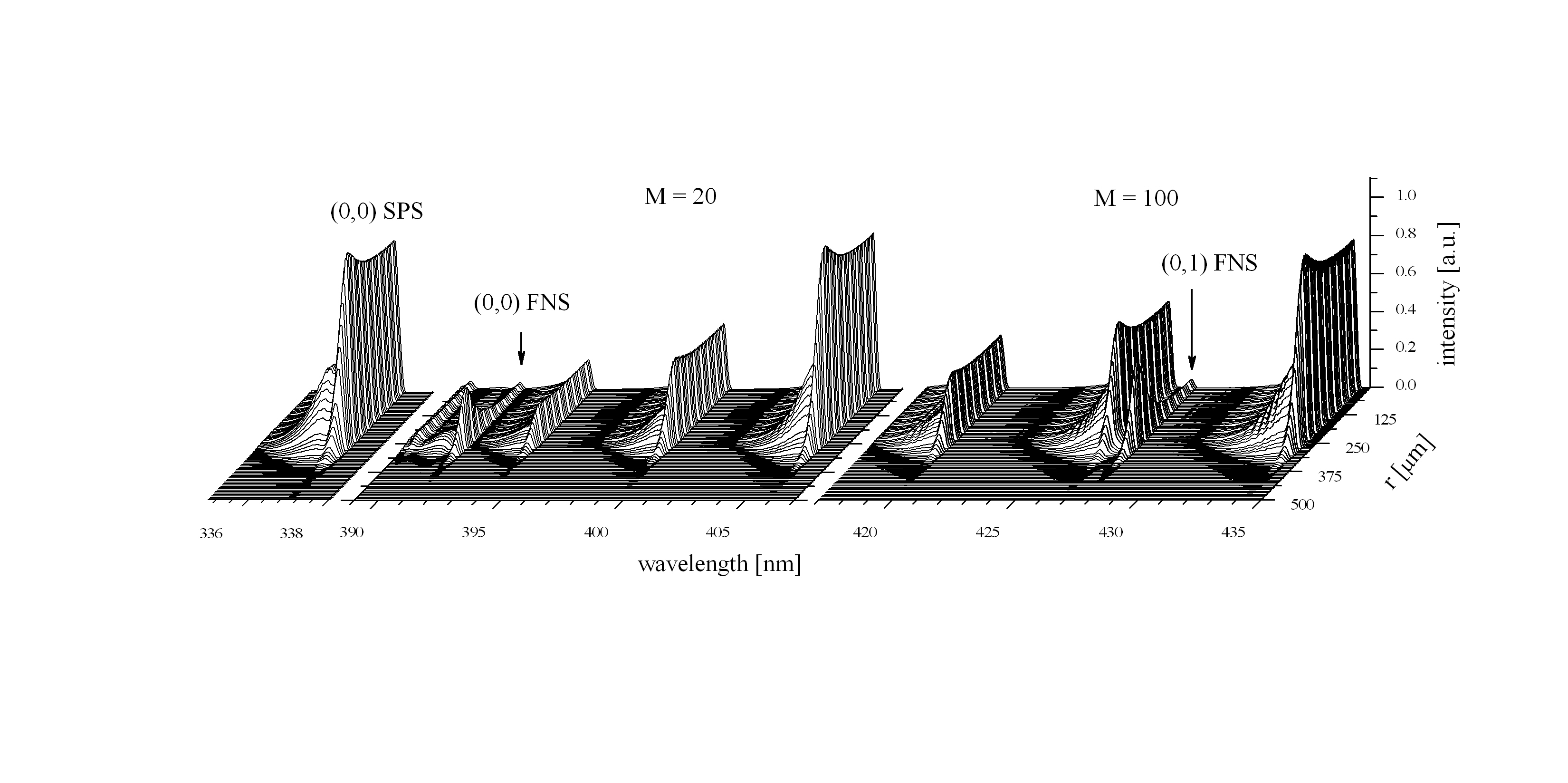}
\end{center}
\caption{Synthetic streamer head emission distribution evaluated along the radius from the data shown in Fig. \ref{head}. Calculated assuming T$_\mathrm{rot}$ = 300$\,$K and using triangular instrumental function (FWHM = 0.2$\,$nm).  Radially resolved spectra simulated at two ($x$ = const.) positions corresponding to maximum axial population (indicated as FNS$_\mathrm{max}$ in Fig.\ref{head}) of the N$_2^+$(B$^2\Sigma)$ state (a) and position slightly shifted ($\Delta x = 112$ $\mu $m, i.e.  approx. 200$\,$ps delay) behind the FNS$_\mathrm{max}$ position (b). The value $M$ denotes the multiplication factor.
}
\label{radial}
\end{figure}

\begin{figure}[h!]
\begin{center}
\includegraphics[width=1\columnwidth]{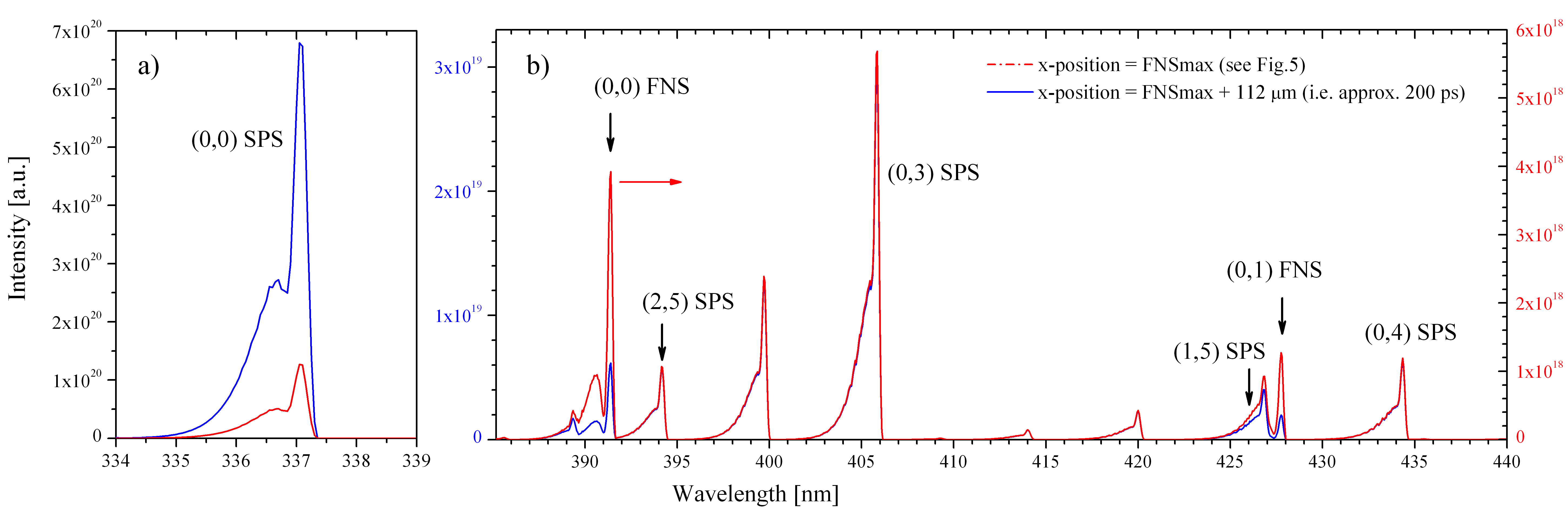}
\end{center}
\caption{
Selected parts of synthetic FNS and SPS spectra occurring in UV (a) and vis (b) ranges  at two different  axial positions averaged over the radius (evaluated from the data shown in Fig. \ref{head}). Calculated assuming T$_\mathrm{rot}$ = 300$\,$K and using triangular instrumental function (FWHM = 0.2$\,$nm). Radially integrated spectra simulated at a position ($x =$ const.) corresponding to maximum axial population of the N$_2^+$(B$^2\Sigma)$ state (FNS$_\mathrm{max}$) and position slightly shifted ($\Delta x = 112\,\mu $m) behind the FNS$_\mathrm{max}$ position (i.e. 200$\,$ps later). 
Note that two different y-axis scales are used in panel (b) to display the two spectra with overlapped (0,3) SPS bands.
}
\label{radialinteger}
\end{figure}

Because obtaining emission data with sufficient spatio-temporal resolution from streamer experiments is very difficult and in most cases impossible due to instrumental limitations, in the next part of this paper we will focus on the radially resolved as well as radially-integrated emission spectra discussing  possible effects connected with determination of basic streamer parameters (emission delays, electric field etc.) from real (experimental) spectra. 
In figure \ref{radial}, the radially resolved spectra of the streamer head are shown for two selected ($x$ = const.) positions behind E$_\mathrm{max}$. The first position (a) coincides with the maximum intensity of the N$_2^+$-FNS system (FNS$_\mathrm{max}$ position) occurring with the shift of 55$\,\mu$m (i.e. approx. 100$\,$ps) behind the E$_\mathrm{max}$ (compare to figures \ref{head} and \ref{sources}). Radial distributions of emission spectra peak at $r = 0$ with the (0,0) FNS band intensity exceeding amplitudes of both (2,5) and (1,4) SPS bands occurring in the 390--400$\,$nm region.
The second position (b) is shifted only by next 112$\,\mu$m (200$\,$ps) with respect to the FNS$_\mathrm{max}$, see the dashed line in Fig. \ref{sources}. 
Radial distributions in the latter case clearly show a shallow dip towards $r = 0$ and a small peak at the periphery of the cylindrically symmetric streamer. The (0,0) FNS band intensity significantly decreases with respect  to SPS bands. 
The main part of the FNS emission is shifted to the side while the SPS emission intensity is nearly constant along the radius with just a small hump occurring at the edge of the streamer channel. The possible uncertainties of the electric field determination via FNS/SPS ratio method by analysing insufficiently resolved (spatially or temporally) streamer head emission are obvious. 
When comparing SPS and FNS amplitudes of radially integrated spectra from two above mentioned axial positions (shown in Fig. \ref{radialinteger}), one can clearly see that when detecting streamer head emission with axial spatial resolution about 0.1$\,$mm, most of the integrated emission comes from regions behind FNS$_\mathrm{max}$ coordinate and therefore FNS/SPS ratio is far from being representative for the electric field estimation at E$_\mathrm{max}$ or even at FNS$_\mathrm{max}$ positions.

The determination of  the three SPS, FNS and FPS axial delay parameters therefore seems to be a crucial step for selecting suitable spatial/temporal resolutions for investigating  fine structure of the streamer head.  
Using SPS-to-FNS delay measurement one can check the relevance of the electric field estimation method. 
The use of the FPS delay parameter together with FNS and SPS ones could be a further improvement of delay-coupled streamer diagnostics. While the FNS and SPS are due to their relatively high excitation thresholds coupled to the streamer head, the FPS is more related to the environment in plasma channel behind the streamer head (see Fig. \ref{domain}). 

Additionally, in synthesised spectra in Fig. \ref{radial} and \ref{radialinteger} another band of the FNS is visible: the (0,1) band. As its intensity is scalable with the usually used (0,0) FNS band only through the ratio of corresponding radiative transition probabilities, it can be easily used for the estimation of the electric field together with neighbouring (1,5) and (0,4) SPS bands which are scalable in the same way 
(considering fixed vibrational distribution of the C$^3\Pi_u$ state) {with the (0,0) SPS band}. This approach has an important advantage because both bands are placed on close wavelengths so there 
is usually no need to make correction for spectral response of the spectrometric system.
It is similar as in \cite{paris2005} where 
the (0,0) FNS and (2,5) SPS bands at 391 nm and 394 nm, respectively were used 
for the same reason as well. 
However in both cases, a possible overlap of FNS band with the tail of SPS band occurring to the next at higher wavelengths has to be carefully evaluated and subtracted (if not negligible).

\section{Summary and conclusion}

In this paper, delays of several tens-of-picoseconds 
for emission maxima of dominant  N$_2$ and N$_2^+$ band systems developing behind the positive streamer head are reported and thoroughly analyzed. 
{In surprising contrast to highly-nonlinear behaviour of streamer events a linear }dependence was found using 2D axi-symmetric simulations as well as 1D analytic models for the emission maxima delays as a function of the streamer radius to velocity ratio $r/v$. 
{We concluded that the coupling of the delay and the $r/v$ parameter represents an 
intrinsic characteristic of the streamer head and we proposed an analytical model for this coupling.} 
A dilatation of these delays of few tens-of-picoseconds during the early-stage streamer evolution 
was observed both experimentally and theoretically with good agreement. 
The SPS delay can reach the value of up to 400$\,$ps at given conditions which can cause an error in the discharge analysis by correlating electrical measurements (current and voltage waveforms for instantaneous power and energy estimation) to the early-stage streamer emission development only.
As shown, the detected emission 
just represents populations of excited states formed behind the running streamer head, giving only delayed information about the streamer head position.
{It is reasonable to expect that similar phenomena occurs in other streamer-mechanism driven discharges, too. Presumably, it can be described by similar analytical approach.}

{For the first time, based on the results of a 2D axi-symmetric model, a spectral representation of the whole streamer head area is visualised. 
It is shown that 
for short delays, the FNS emission peaks at the axis of the streamer while several tens of picoseconds later more complex FNS intensity distributions appear.} 
Also, by understanding the spectra structure in the streamer head, one can assess that the radial averaging of the streamer emission will cause smaller distortion of the further processed signal as the axial signal integration. 

Finally, we suggest to consider these delays in advanced emission-based streamer diagnostics. 
Analysing accurately this effect for selected case {and using the analytical expression for the delay parameter derived in this article, we see possibilities to assess more easily the basic plasma parameters of the propagating streamer, based on the measurements of these delays and other macroscopic streamer parameters only.
 }

\section*{Acknowledgments}
One of authors (TH) is grateful to Ronny Brandenburg (INP Greifswald) for supporting this work. 
TH was supported by the Federal German Ministry of Education and partly by European Science Foundation exchange grant no. 4219 within the TEA-IS network.
ZB acknowledges support by project CZ.1.05/2.1.00/03.0086 funded by European Regional Development. MS acknowledges the support of the Czech Science Foundation (contract P205/12/1709).


\end{document}